\documentclass[twocolumn,secnumarabic,amssymb, nobibnotes, aps, prl, superscriptaddress]{revtex4-1}

\newcommand{\braket}[1]{\langle{}#1\rangle}

\usepackage{graphicx}
\usepackage{units}
\usepackage{color}
\usepackage{float}

\begin{document}

\title{Superconducting Nanowire Single Photon Detector on Diamond}%

\author{Haig A. Atikian}
\affiliation{Harvard University, School of Engineering and Applied Sciences, 33 Oxford Street, Cambridge, USA}
\author{Amin Eftekharian}
\affiliation{University of Waterloo, 200 University Ave West, Waterloo, ON, Canada, N2L 3G1}
\affiliation{Institute for Quantum Computing, University of Waterloo, 200 University Ave West, Waterloo, ON, Canada, N2L 3G1}
\author{A. Jafari Salim}
\affiliation{University of Waterloo, 200 University Ave West, Waterloo, ON, Canada, N2L 3G1}
\affiliation{Institute for Quantum Computing, University of Waterloo, 200 University Ave West, Waterloo, ON, Canada, N2L 3G1}
\author{Michael J. Burek}
\affiliation{Harvard University, School of Engineering and Applied Sciences, 33 Oxford Street, Cambridge, USA}
\author{Jennifer T. Choy}
\affiliation{Harvard University, School of Engineering and Applied Sciences, 33 Oxford Street, Cambridge, USA}
\author{A. Hamed Majedi}
\affiliation{University of Waterloo, 200 University Ave West, Waterloo, ON, Canada, N2L 3G1}
\affiliation{Institute for Quantum Computing, University of Waterloo, 200 University Ave West, Waterloo, ON, Canada, N2L 3G1}
\author{Marko Lon$\mathrm{\breve{c}}$ar}\email{loncar@seas.harvard.edu}
\affiliation{Harvard University, School of Engineering and Applied Sciences, 33 Oxford Street, Cambridge, USA}

\begin{abstract}
Superconducting nanowire single photon detectors (SNSPDs) are fabricated directly on diamond substrates and their optical and electrical properties are characterized.  Dark count performance and photon count rates are measured at varying temperatures for 1310nm and 632nm photons. The procedure to prepare diamond substrate surfaces suitable for the deposition and patterning of thin film superconducting layers is reported.  Using this approach, diamond substrates with less than 300pm RMS surface roughness are obtained. 
\end{abstract}

\maketitle

Diamond has recently gained significant interest as a promising platform for on-chip high-performance photonic devices \cite{Aharonovich2011, Hausmann2012PSS, Loncar2013}. Diamond exhibits many favorable material properties such as a high refractive index (n=2.4), wide bandgap (5.5eV), and a large optical transmission range from the UV to the mid infrared.  Diamond is also host to numerous defect color centers such as the nitrogen-vancancy (NV) center, that can be utilized as an optically addressable spin based memory, particularly interesting in the field of quantum information processing \cite{Neumann2008, O'Brien2007}. In addition, diamond has a relatively large Kerr non-linearity \cite{Boyd2008} ($n_2=\unit[1.3\cdot10^{-19}]{m^2/W}$) making it an attractive platform for on-chip nonlinear optics in the visible and infrared wavelengths \cite{Hausmann2013}.  An exciting application for utilizing this diamond non-linearity could allow for frequency conversion of photons generated by color centers in diamond, which typically emit in the visible range, to the telecom wavelengths \cite{Raymer2012}.  This could enable transmission of quantum information and distribution of quantum entanglement \cite {Togan2010, Bernien2013} over long distances, for the realization of quantum repeaters. Such an integrated diamond quantum photonics platform would benefit from the realization of high performance single photon detectors, with broadband photon sensitivity, that are integrated directly on the same diamond chip.

Superconducting nanowire single photon detectors (SNSPDs) outperform other single photon detector technologies on several merits such as quantum efficiency {\cite{Marsili2013}, timing jitter, dark count rates, and broad spectral sensitivity \cite{Hadfield2009, Akhlaghi2012}.  SNSPDs typically consist of narrow width nanowires patterned into an ultrathin (4nm to 8nm) superconducting film, commonly made from niobium nitride or some derivative of \cite{Dorenbos2008}. The nanowires are current biased close to the critical current of the superconductor.  When an incident photon is absorbed by the wire, a small resistive hotspot is formed generating a voltage pulse that can be subsequently amplified and measured \cite{Goltsman2001}.  Since the performance of SNSPDs is critically dependent on nanowire structural uniformity, it is crucial to have them deposited on smooth substrate surfaces to avoid constrictions that can have detrimental effects on detection efficiency\cite{Kerman2001}.  

In this letter, we report NbTiN superconducting nanowires deposited directly on diamond substrates that exhibit promising single photon sensitivity.  Specifically, we provide details of the fabrication procedure developed, resulting in smooth diamond surfaces suitable for deposition of high-quality superconducting thin films.  Furthermore, we show experimental data on the temperature performance of the SNSPDs on diamond substrates irradiated by 632nm and 1310nm photons. The performance of the devices is comparable to the performance of SNSPDs fabricated on conventional substrates such as sapphire, MgO, and silicon \cite{Hadfield2009}. 

\begin{figure*}
	\centering
	\includegraphics[width=0.8\textwidth]{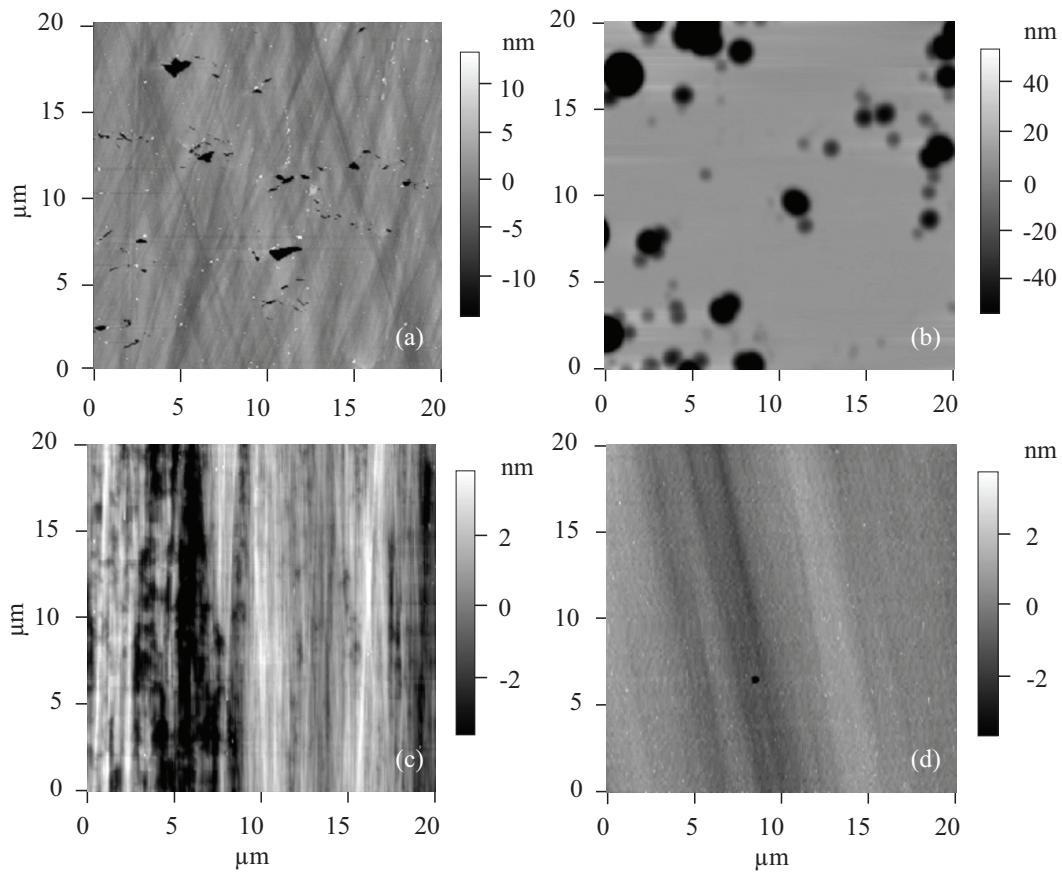}
	\caption{(a) AFM image of a type IIa diamond surface from Element Six after acid cleaning in a 1:1:1 refluxing mixture of sulfuric, nitric and perchloric acid with an initial RMS surface roughness of $\thicksim$30nm. (b) AFM image of a IIa diamond surface after Ar/Cl$_{2}$ and O$_{2}$ RIE pre-etch. (c) AFM image of a IIa diamond surface after mechanical polishing with RMS surface roughness of 1-2nm. (d) AFM image of the final type IIa diamond surface obtained after another Ar/Cl$_{2}$ and O$_{2}$ RIE with an RMS surface roughness of $\thicksim$300pm.}
	\label{fig:AFM}
\end{figure*}

We use type IIa single crystal diamond substrates cut in the \{100\} crystallographic orientation, obtained from Element Six, which were grown using microwave assisted CVD. The "as received" diamond is polished along random crystal directions leading to the rough surface seen in Fig.\ref{fig:AFM}(a).  Diamond polishing involves sliding the diamond specimen on a cast iron wheel called a 'sciafe' charged with diamond powder at high pressures and high velocities. A terminology used in the polishing industry and scientific literature refers to the direction of the highest polishing wear rate (in the case of our IIa diamond samples the $\braket{100}$ direction on the \{100\} crystal plane), as the "easy direction" and the direction with the lowest polishing wear rate (in our samples the $\braket{110}$ or $\braket{111}$ direction in the \{100\} crystal plane) the "hard direction" . Work by Grillo and Field \cite{Grillo1997} showed that when polishing \{100\} diamonds in the $\braket{110}$ direction, or the "hard direction", the polishing debris consists of fragments of diamonds, suggesting the material is being removed by mechanically chipping the surface.  This is clearly seen in Fig.\ref{fig:AFM}(a) where large chips, several microns in size, are seen on the diamond surface.  They also showed that when polishing \{100\} diamond in the $\braket{100}$ direction, or the "easy direction", the wear debris mostly consisted of less dense forms of sp$^2$ carbon material, where material can be removed more efficiently without chipping. The randomly oriented striations seen on the diamond surface also suggests how the polishing was done without aligning the diamond to its "easy direction" and leading to the hatch patterned polishing lines seen in Fig.\ref{fig:AFM}(a).  

Since the surface of these diamonds are typically damaged and highly strained due to the mechanical nature of cutting and polishing, the surface must be prepared before deposition of the superconducting thin film.  The first step is an acid clean to remove the surface contaminants from the diamond. The diamonds are cleaned in a 1:1:1 refluxing mixture of sulfuric, nitric, and perchloric acid for approximately 1 hour.  Fig.\ref{fig:AFM}(a) shows an atomic force microscopy (AFM) image of the diamond surface, as received from Element Six, after the acid clean. It exhibits an RMS surface roughness of $\thicksim$30nm, with some clear chipped surface defects. This AFM image is corroborated by previous works done on diamond polishing \cite{Grillo2000}.  This surface is clearly not suitable to deposit thin film superconductors, further steps are required to prepare an appropriate diamond surface. 

\begin{figure*}
	\centering
	\includegraphics{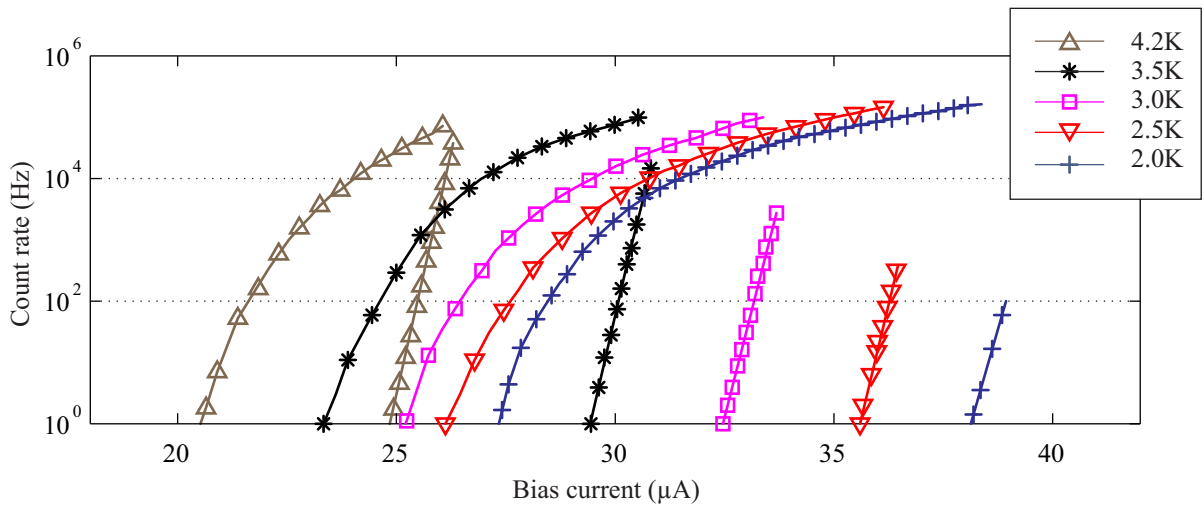}
	\caption{Count rate versus bias current for a 100nm wide 6nm thick nanowire, for temperatures from 2K to 4.2K. Each temperature data set has two curves represented by the same color and symbol. For each temperature, the straight line to the right represents the dark counts with no laser irradiation, while the left curve represents the count rate to an attenuated 1310nm laser source.}
	\label{fig:CRvsIc}
\end{figure*}

The next step is to remove several microns from the surface layer of the diamond to be clear of any mechanically strained areas caused by the polishing process. This is achieved by using a two step reactive ion etching (RIE) pre-etch process.  We first use a 30min Ar/Cl$_{2}$ etch, followed by another $30$min of O$_{2}$ etch in the UNAXIS Shuttleline inductively coupled plasma (ICP) RIE. The resulting diamond surface can be seen in Fig.\ref{fig:AFM}(b). Many rounded deep etch pits are formed in the diamond on the order of 100's of nanometers.  These etch pits are believed to be formed in the vicinity of the defects on the original diamond surface created by polishing.  The damaged areas etch at a different rate than the more dense diamond regions creating this cratered surface profile.

Following the RIE step, the diamond is re-polished along its "easy direction" to prepare a smooth surface. This second mechanical polishing step reduces the RMS surface roughness to the order of 1-2 nanometers as seen in Fig.\ref{fig:AFM}(c). The polishing striations seen in this figure are all aligned in the same direction, which is a strong indication that the polishing was carefully performed along the "easy direction". Since the thickness of our superconducting nanowires is typically around 5nm, having a surface roughness in this range would still be detrimental to device performance.  Therefore, we follow this step with another acid clean with the same conditions as before to remove surface contaminants from the polishing process, and perform another RIE etch again with 30min of Ar/Cl$_{2}$ and 30min of O$_{2}$.  This RIE etch is used to access a portion of the diamond that is clear of mechanical stress created by polishing, and to further average out the surface roughness via the ICP \cite{Lee2006}. The final surface is seen in Fig.\ref{fig:AFM}(d) where the RMS surface roughness is $\thicksim$150-300pm, depending on the sample, for an area without any of the small circular defects.  A large scan of the diamond chip showed that these small circular pit defects are sparsely present throughout the sample and typically 5nm in depth.  The probability of patterning a nanowire that does not intersect with one of these defects is quite high, and necessary to make a good device.  Similar surface roughness was achieved in Ib diamonds using the procedure outlined above. 

A NbTiN layer is deposited directly on the diamond surface by reactive sputtering using a Nb-Ti alloy target and a dc magnetron source. The Ar/N$_{2}$ partial pressure ratio and total pressure have been optimized to maximize critical temperature, and the substrates are held at ambient temperature. The critical temperature was measured to be 8.5K, taken at the midpoint of the superconducting transition. Hydrogen silsesquioxane (HSQ) resist is used to pattern the nanowire structures using 125keV electron-beam lithography.  The resist is developed in a 25$\%$ tetra-methyl ammonium hydroxide solution, and the nanowire pattern is transferred into the film using ion beam milling with Argon gas.

A straight nanowire 10$\mathrm{\mu{}m}$ long and 100nm wide is connected to larger gold contact pads with a gradual taper to avoid any current crowding effects caused by bends and abrupt width variations \cite{Clem2011}. The measurements were performed in a dipstick probe immersed in a Dewar of liquid helium.  Each nanowire is connected to two series 470nH inductors (placed next to the sample) and to a room temperature bias-T via 50$\Omega$ coax cables. The inductors are utilized such that the electrical time constant is larger than the thermal time constant of the detector to avoid latching, ensuring the peak measured device current is
the "experimental" critical current \cite{Kerman2009}. A computer controlled source meter (Keithley 2400) is connected to the DC port of the bias-T via a low pass filter.  The high frequency pulses generated at the nanowire are amplified at room temperature and monitored on a high speed oscilloscope and programmable counter. A single mode fiber is placed several centimeters away from the sample, uniformly irradiating the sample with photons from an attenuated laser source. 

\begin{figure}
	\centering
	\includegraphics{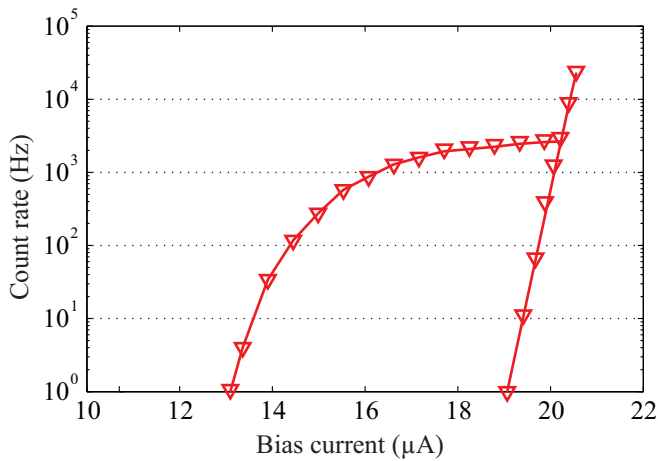}
	\caption{Count rate versus bias current at 4.2K for a 100nm wide 6nm thick nanowire irradiated by 632nm photons. The straight line to the right represents the dark counts with no laser irradiation, while the left curve represents the count rate to an attenuated 632nm laser source.}
	\label{fig:red}
\end{figure}

Figure \ref{fig:CRvsIc} illustrates the photon counting performance of a 100nm wide, 6nm thick nanowire on a diamond substrate irradiated by 1310nm photons at varying temperatures. The impact of hotspot formation on the current distribution of the nanowire depends on the relative size of the generated hotspot area to the transverse dimension of the nanowire. In experiment, to have a device sensitive to single photons, the bias current is set close to the depairing current of the superconducting material. When the formation of the initial hotspot reaches its maximum effective size \cite{Zotova2012,Eftekharian2013}, the local perturbation of current density is significant enough that a hotspot-driven photon-count event is triggered. For a bias current equal to or greater than this minimum value, all the photon absorptions are registered as photon-counts and hence the detector operates independent of the bias current. This behavior refers to the saturated area of quantum efficiency versus bias current observed in Fig.\ref{fig:CRvsIc} and \ref{fig:red}. In SNSPDs, even when the input optical terminal is fully blocked, randomly distributed signals are generated at the output electrical terminal of the detector for a particular current bias condition. Regardless of the nature of these thermal fluctuations, the rate always obeys the Kramers formula \cite{Bulaevskii2012,Bartolf2010,Yamashita2009} which predicts an exponential dependence of the rate to the height of the potential barrier faced by these thermal fluctuations. Since this potential barrier is current dependent, a power law dependence is predicted for the dark count versus bias current measurement as observed in Fig.\ref{fig:CRvsIc}. These curves are indicative of a nanowire with uniform superconductive properties and no appreciable constrictions which have a drastic effect on photon counting performance\cite{Kerman2001}. 

Fig.\ref{fig:red} illustrates the photon counting performance of a 100nm wide, 6nm thick nanowire on a diamond substrate irradiated by 632nm photons at 4.2K. The motivation is to show the nanowire's ability to detect photons emitted by NV centers in diamond with a zero-phonon line at 637 nm and phonon sideband from 640 to 780 nm.  In this case of higher energy photons, the photo-generated hotspot has a greater effective size and can create a photon detection event at lower bias currents. Consequently, the count rate to 632nm photons reaches a saturation level well before it intersects the dark count line even at 4.2K. For NV applications these detectors can be biased at an operating point far from the critical current. Looking at Fig.\ref{fig:red}, there exists a wide window where the detector can operate with practically zero dark counts, and a photon count rate in the saturated region with little influence to changes in bias current. 

In conclusion, we have presented a method to prepare diamond surfaces with sub 300pm RMS surface roughness suitable to deposit long superconducting nanowires for the purpose of single photon detection. We have also shown that the added challenges diamond presents to prepare its surface to a fine roughness does not hinder the performance of contemporary SNSPDs, and allow for easy integration of SNSPDs on nanoscale photonic devices carved out of single crystal diamond as the ones seen in \cite{Burek2012}.  This approach is suitable for realization of a fully integrated quantum photonics platform based on diamond, that leverages diamond's color centers as sources, and SNSPDs as detectors of single photons. 

This work was performed in part at the Center for Nanoscale Systems (CNS), a member of the National Nanotechnology Infrastructure Network (NNIN), which is supported by the National Science Foundation under NSF award no. ECS-0335765. CNS is part of Harvard University. We acknowledge the financial support of NSERC, OCE, and IQC. This work was supported in part by the DARPA QuINESS Program and the Center for Excitonics, an Energy Frontier Research Center funded by the U.S. Department of Energy, Office of Science, Office of Basic Energy Sciences under Award Number DE-SC0001088. The authors would like to acknowledge Robin Cantor for helpful comments. The authors thank Daniel Twitchen from Element Six for support with diamond samples.

\bibliographystyle{apsrev}
\bibliography{dsspd}

\end{document}